\documentclass{PoS}
\usepackage{braket,amsmath}
\renewcommand{\d}{\ensuremath{\mathrm{d}}}

\newcommand{\p}{\partial}

\newcommand{\occ}{\overline{c}}

\title{Aspects of the Gribov-Zwanziger framework}

\ShortTitle{The GZ framework}

\author{\speaker{David Dudal}, Nele Vandersickel, Henri Verschelde \\
        Ghent University, Department of Mathematical Physics and Astronomy \\
        Krijgslaan 281-S9, 9000 Gent,Belgium\\
        E-mail: \email{nele.vandersickel@ugent.be} ,\email{ david.dudal@ugent.be}, \email{henri.verschelde@ugent.be }}

\author{Silvio P. Sorella\\
        Departamento de F\'{\i }sica Te\'{o}rica, Instituto de F\'{\i }sica, UERJ - Universidade do Estado do Rio de Janeiro\\
        Rua S\~{a}o Francisco Xavier 524, 20550-013 Maracan\~{a}, Rio de Janeiro, Brasil\\
        E-mail: \email{sorella@uerj.br}
}

\abstract{The existence of gauge (Gribov) copies disturbs the usual Faddeev-Popov
quantization procedure in the Landau gauge. It is a very hard job to treat these in the continuum,
even in a partial manner. A decent way to do so was worked out by Gribov, and later on by
Zwanziger. The final point was a renormalizable action (the Gribov-Zwanziger action), implementing the
restriction of the path integration to the so-called Gribov region, which is free of a subset of gauge
copies, but not of all copies. Till recently, everybody agreed upon the fact that the restriction to the Gribov
region implied a infrared enhanced ghost, and vanishing zero momentum gluon propagator. We
discuss how the Gribov-Zwanziger action naturally leads to the existence of vacuum condensates of dimension
two. As it is very common, such condensates can seriously alter the dynamics. In particular, the
Gribov-Zwanziger condensates give rise to a gluon propagator with a finite but nonvanishing zero momentum limit, and reconstitute
a nonenhanced ghost. We call this the refined Gribov-Zwanziger
 framework. The predictions are
in qualitative agreement with most recent lattice simulations, and certain solutions of the Schwinger-Dyson
equations. A crucial feature of the Gribov-Zwanziger framework is the soft (controllable) breaking of the BRST
symmetry. We also point out that imposing the Kugo-Ojima confinement criterion on the Faddeev-Popov
theory as a boundary condition from the beginning leads to the same partition function as of Gribov-Zwanziger,
with associated BRST symmetry breaking. This clouds the interpretation of the Kugo-Ojima criterion in se.}

\FullConference{International Workshop on QCD Green's Functions, Confinement, and Phenomenology - QCD-TNT09\\
		 September 07 - 11 2009\\
		 ECT Trento, Italy}

\begin{document}

\section{The Gribov-Zwanziger framework}
As it is well-known, the classical $SU(N)$ Yang-Mills action,
\begin{eqnarray}
S_{YM} &=& \frac{1}{4}\int \d^4 x F_{\mu\nu}^a F_{\mu\nu}^a,
\end{eqnarray}
displays an enormous invariance w.r.t.
\begin{eqnarray}\label{gi}
A_\mu^a\to A_\mu^a + D_\mu^{ab}\omega^b
\end{eqnarray}
with $\omega^a$ arbitrary (infinitesimal) functions\footnote{We shall not speak about ``large'' gauge transformations which are not connected to the unity.}.  In order to quantize this gauge theory, one needs in principle to select a single gauge representant $A_\mu^a$ by fixing the gauge. A popular choice is to restrict to the transverse degrees of freedom by imposing the Landau gauge, $\p_\mu A_\mu^a=0$.

After some nice machinery at the level of the path integral, one ends up with the Faddeev-Popov gauge fixed action in the Landau gauge.
\begin{eqnarray}\label{fp}
S_{YM+FP} &=& \frac{1}{4}\int \d^4 x F_{\mu\nu}^a F_{\mu\nu}^a+\int \d^{4}x\;\left( b^{a}\partial_\mu A_\mu^{a}+\overline{c}^{a}\partial _{\mu } D_{\mu }^{ab}c^b \right),
\end{eqnarray}
which comes from the partition function $Z$ by
\begin{equation}
    Z=\int [\d A] \delta(\p A) \det M e^{-S_{YM}}=\int [\d A][\d b][\d c][\d\occ]  e^{-S_{YM+FP}}.
\end{equation}
Indeed, setting $b^a=iB^a$, one finds the functional analog of the Fourier representation of a $\delta$-function. The accompanying Jacobian $\det M$ is lifted into the exponential using the ghost fields $c^a$ and  $\occ^a$.  We introduced the Faddeev-Popov operator as
\begin{equation}
    M^{ab}=-\p_\mu D_\mu^{ab}=-\p_\mu \left( \p_{\mu} \delta^{ab} + g f^{acb} A^c_{\mu} \right).
\end{equation}
Clearly, the local gauge invariance expressed by \eqref{gi} is lost. It is however replaced by the BRST invariance \cite{Becchi:1975nq,Tyutin},
\begin{align*}
sA_{\mu }^{a} &=-\left( D_{\mu }c\right) ^{a}\;, & sc^{a} &=\frac{1}{2}gf^{abc}c^{b}c^{c}\;,   & s\overline{c}^{a} &=b^{a}\;,&   sb^{a}&=0,
\end{align*}
with nilpotent generator $s$, $s^2=0$. This BRST symmetry is a cornerstone of (perturbative) gauge theories.

In the above construction, it was silently assumed that there is only one solution $A_\mu^a$ on the gauge orbit of an arbitrary gauge field that obeys the Landau gauge, i.e. $\p_\mu A_\mu=0$. Let us consider a gauge equivalent field $A_\mu'$,
\begin{eqnarray}
A_\mu^a\,'= A_\mu^a + D_\mu^{ab}\omega^b,
\end{eqnarray}
then this $A_\mu'$ could also obey the Landau gauge condition, $\p_\mu A_\mu'=0$, if
\begin{equation}
    M^{ab}\omega^b=0,
\end{equation}
i.e.~if the Faddeev-Popov operator has a zero mode. It was the insight of Gribov to show that such zero modes do exist, hence we encounter the problem of gauge copies \cite{Gribov:1977wm}. This of course poses a serious problem for the Faddeev-Popov quantization procedure, where it was e.g. assumed that $\det M\geq 0$, but since $M$ has zero modes, it can switch sign\footnote{We recall that the $\delta$-function is accompanied by the absolute value of its corresponding Jacobian. It is however unknown how to lift $|\det M|$ into the exponential.}. For perturbation theory, these copies are pretty harmless, as discussed in e.g.~\cite{Frishman:1978ke}. However, when one is interested in a region where nonperturbative physics could enter the game, one needs to take care of these copies. Therefore, it would be interesting to improve upon the original Faddeev-Popov procedure.

This has been worked out by Gribov in his seminal paper in a steepest descent approximation. We quote here his result, which is an improved version of the Landau gauge fixed partition function, reading
\begin{eqnarray}\label{gribov}
Z=\int [\d\Phi] e^{-S_{YM+FP}+\gamma^4 \int \d^4x A\frac{1}{\p^2}A}.
\end{eqnarray}
This was achieved by writing the inverse Faddeev-Popov operator (with an external gauge field) as
\begin{equation}\label{nop}
    M^{-1}=(-\p D)^{-1}=\frac{1}{k^2(1-\sigma(k^2,A))}
\end{equation}
and inserting into the path integral a step function $\theta(1-\sigma(0,A))$, which ensures that are are only integrating in $A$-space over configurations wherefore $M$ is positive. The region
\begin{eqnarray*}
\Omega &\equiv &\{ A^a_{\mu}, \, \p_{\mu} A^a_{\mu}=0, \, M^{ab}  >0  \}
\end{eqnarray*}
is known as the Gribov region, and it displays many nice properties, summarized in \cite{Sobreiro:2005ec}. For example, $\Omega$ is convex, bounded in all directions in field space, and it has the crucial property that each gauge orbit intersects with $\Omega$.

After a few steps, one then arrives at \eqref{gribov}, whereby $\gamma^4$ is a thermodynamic mass parameter, fixed by
\begin{equation}\label{gap}
\frac{3}{4}g^2N\int\frac{\d^4q}{(2\pi)^4}\frac{1}{q^4+\gamma^4}=1.
\end{equation}
We notice here that this gap equation is ill-defined as the integral diverges. The condition
\begin{equation}\label{nop2}
\sigma(0,A)\leq 1
 \end{equation}
is known as the Gribov no-pole condition. We notice here that if $A$ becomes a quantum field and is integrated over, the expectation value of the operator \eqref{nop} corresponds exactly to the ghost propagator.

By implementing the restriction to the Gribov horizon, one already avoids copies which are infinitesimally related to each other. Unfortunately, there are still other copies left inside $\Omega$, see e.g. \cite{vanBaal:1991zw}. A further restriction would be necessary, but it is unclear how this could be achieved in an analytical fashion.

Having an improved partition function at our disposal, one has the ideal tool to study the theory. In particular, one can look at the elementary gluon and ghost propagator. For the gluon propagator, one finds at tree level\footnote{We set $\lambda^4=2g^2N\gamma^4$.}
\begin{equation}\label{gl1}
    \braket{A_\mu^a A_\nu^a}_p=\delta^{ab}D(p^2)\left(\delta_{\mu\nu}-\frac{p_\mu p_\nu}{p^2}\right),\qquad D(p^2)=\frac{p^2}{p^4+\lambda^4},
\end{equation}
while for the ghost,
\begin{equation}\label{gh1}
    \braket{c^a \occ^b}_p=\delta^{ab}G(p^2),\qquad G(p^2)=\frac{\delta^{ab}}{p^2}\frac{1}{1-\sigma(p^2)},
\end{equation}
this at one loop. Implementing the gap equation leads to $\sigma(0)=1$, thus the ghost propagator exhibits an infrared enhancement. This should not come as a big surprise, since the gap equation \eqref{gap} is the quantum version of the no-pole condition \eqref{nop2}. At $p^2=0$, we reach the boundary value $\sigma(0)=1$, which corresponds to the boundary (horizon) $\p\Omega$ of the Gribov region $\Omega$. For the gluon propagator, we notice an infrared suppression, in particular $D(0)=0$.

So far, everything we have mentioned is restricted to the lowest (quadratic) order. After a tour-de-force, Zwanziger was able to extend Gribov's result to all orders \cite{Zwanziger:1989mf,Zwanziger:1992qr}, giving the following action
\begin{eqnarray}\label{gzz}
S_h&=& S_{YM+FP}  + \gamma^4 \int \d^4 x\; h(x),
\end{eqnarray}
with the horizon function
\begin{eqnarray}\label{hcond}
 h(x) &=& g^2 f^{abc} A^b_{\mu} \left(M^{-1}\right)^{ad} f^{dec} A^e_{\mu}.
\end{eqnarray}
The parameter $\gamma$ is now fixed by demanding that
\begin{eqnarray}
\braket{h(x)} &=& d (N^2 -1),
\end{eqnarray}
the latter being known as the horizon condition. It can be easily checked that the Zwanziger results reduce to those of Gribov at lowest order. We draw attention to the fact that the action is a nonlocal one. As nonlocal quantum field theories are hard to handle and interpret, and almost all available tools are intended for local quantum field theories, this looks troublesome. Fortunately, by introducing a suitable set of extra fields, we can reformulate \eqref{gzz} as
\begin{eqnarray}
S_{GZ} &=&S_{YM+FP} + S_{h},
\end{eqnarray}
with
\begin{eqnarray}\label{gz1}
S_{h} &=&
\int \d^{4}x\Bigl( \overline{\varphi }_{\mu }^{ac}\partial _{\nu}\left(\partial _{\nu }\varphi _{\mu }^{ac}+gf^{abm}A_{\nu }^{b}\varphi _{\mu}^{mc}\right)
-\overline{\omega }_{\mu }^{ac}\partial _{\nu }\left( \partial_{\nu }\omega _{\mu }^{ac}+gf^{abm}A_{\nu }^{b}\omega _{\mu }^{mc}\right)
\nonumber  \\
&&\!\!-g\left( \partial _{\nu }\overline{\omega }_{\mu}^{ac}\right) f^{abm}\left( D_{\nu }c\right) ^{b}\varphi _{\mu}^{mc}   -\gamma ^{2}g\left(f^{abc}A_{\mu }^{a}\varphi _{\mu }^{bc}+f^{abc}A_{\mu}^{a}\overline{\varphi }_{\mu }^{bc} + \frac{4}{g}\left(N^{2}-1\right) \gamma^{2} \right)\Bigr),
\end{eqnarray}
which is known as the Gribov-Zwanziger action. The horizon condition \eqref{hcond} turns out to be equivalent to
\begin{eqnarray}\label{hcond2}
\frac{\p \Gamma}{\p \gamma^2}&=&0 \Leftrightarrow \left\langle gf^{abc} A_\mu^a (\varphi+\overline{\varphi})_{\mu}^{bc} \right\rangle=2d(N^2-1)\gamma^2,
\end{eqnarray}
where the artificial solution $\gamma^2=0$ needs to be discarded. We see that the horizon condition corresponds to a certain nontrivial $d=2$ condensate.

Having constructed a local action implementing the restriction to $\Omega$, one can really start to study the theory. In particular, we can mention that $S_{GZ}$ defines a renormalizable theory, which is a highly nontrivial issue \cite{Zwanziger:1992qr,Maggiore:1993wq,Dudal:2005na}. In particular, we draw attention to the fact that the factor in front of the pure vacuum term in \eqref{gz1} turns out to be uniquely fixed by the renormalization. If this would be not the case, as it is in general the case \cite{Knecht:2001cc}, it would be impossible to speak about the renormalized version of the horizon condition \eqref{hcond2}, and thus of the gap equation \eqref{gap}. We have found this to be a truly remarkable feature of the Gribov-Zwanziger framework. For the propagators, one basically finds back the Gribov predictions at higher order too, i.e.~an infrared suppressed gluon propagator and a enhanced ghost propagator. This can be proven using the Ward identities of the theory as done in \cite{Zwanziger:1992qr}, while it was also explicitly verified in works as \cite{Ford:2009ar,Gracey:2009mj}.

\section{Fate of the BRST symmetry in the Gribov-Zwanziger framework}
Let us first mention that for $\gamma=0$, the physics described by the action \eqref{gz1} is completely equivalent to that of the original Faddeev-Popov theory. This can be appreciated since there exists a natural extension of the original BRST transformation to these fields, by setting
\begin{align*}
s\varphi _{\mu}^{ab} &=\omega _{\mu}^{ab}\;,&s\omega _{\mu}^{ab}&=0\;, & s\overline{\omega }_{\mu}^{ab} &=\overline{\varphi } _{\mu}^{ab}\;,& s \overline{\varphi } _{\mu}^{ab}&=0\;, \end{align*}
The action \eqref{gz1} is then an $s$-exact piece, $S_h=s(\ldots)$, while the extra fields come as $s$-doublets, hence they are physically trivial \cite{Piguet:1995er}. In the case of interest, $\gamma\neq0$, and we arrive at a more peculiar conclusion. We find
\begin{eqnarray}
   sS_{GZ} &=& g \gamma^2 \int \d^4 x f^{abc} \left( A^a_{\mu} \omega^{bc}_\mu -
 \left(D_{\mu}^{am} c^m\right)\left( \overline{\varphi}^{bc}_\mu + \varphi^{bc}_{\mu}\right)  \right) ~=~ \Delta_\gamma,
\end{eqnarray}
or we observe a soft BRST breaking. The breaking is called soft, as it is only quadratic in the fields. Therefore, it is perfectly under control at the quantum level, and one can even write down the corresponding softly broken Slavnov-Taylor identity \cite{Dudal:2008sp},
\begin{eqnarray}\label{sti}
\int \d^{4}x\left( \frac{\delta \Gamma}{\delta K_{\mu }^{a}}\frac{\delta \Gamma }{\delta A_{\mu}^{a}}+\frac{\delta \Gamma }{\delta L^{a}}\frac{\delta \Gamma}{\delta c^{a}}+b^{a}\frac{\delta \Gamma}{\delta \overline{c}^{a}}+\overline{\varphi }_{i}^{a}\frac{\delta \Gamma }{\delta \overline{\omega }_{i}^{a}}+\omega _{i}^{a}\frac{\delta \Gamma }{\delta \varphi _{i}^{a}}\right)
&=& -\left[ \Delta_{\gamma} \cdot \Gamma \right],
\label{stdelta}
\end{eqnarray}
whereby  $\left[ \Delta_{\gamma} \cdot
\Gamma \right]$ corresponds to insertion of the breaking operator
 $\Delta_{\gamma}$. A nice application is that this allows to prove that the gluon self energy is no longer transverse. This was explicitly checked to two loops in \cite{Gracey:2009mj}.

At the level of renormalizability, there is thus no problem associated with the softly broken BRST. However, as it is well know, the BRST also plays a crucial role in establishing the unitarity of gauge theories \cite{Kugo:1979gm}, including the definition of the physical states, which a fortiori must have a positive norm. Let us therefore first give a quick look at the usual Faddeev-Popov case.

\section{The Kugo-Ojima analysis}
In \cite{Kugo:1979gm}, building on work \cite{Curci:1976yb}, a beautiful discussion of the unitarity issue based on the nilpotent BRST charge was given. When physical states $\left|\psi_p\right\rangle$ are defined as those states belonging to the BRST cohomology\footnote{It is of no use to keep the BRST-exact states, $Q_{BRST}\left|\ldots\right\rangle$, as these are trivially annihilated by $Q_{BRST}$, moreover, they have zero norm.}, $\left|\psi_p\right\rangle \in \textrm{cohom}(Q_{BRST})=\frac{\mathrm{Ker} Q_{BRST}}{\mathrm{Im} Q_{BRST}}$, one can show that only the 2 transverse gluon polarizations are physical; the other two cancel with the (anti)ghost particles. This is a generic feature, mainly based on purely algebraic properties,  stating that unphysical particles always decouple in quartets (the members of a quartet are related to each other by the BRST and ghost charge). Roughly speaking, this property remains preserved under time evolution, as the BRST charge commutes with the Hamiltonian as a symmetry generator. Said otherwise, it allows one to define a unitary ${\cal S}$-matrix.

Kugo-Ojima continued their analysis, and also considered the global color charge, which can be written as\footnote{We do not pay attention here to working in Minkowksi or Euclidean space. The ideas thereafter will be clear.}
\begin{eqnarray}\label{color}
Q^a = \int \d^3x \p^i F_{i0}^a+\int \d^3x\{Q_{BRST},D_0 \overline{c}^a\}.
\end{eqnarray}
Assuming that gluons are infrared suppressed in the sense that the first volume-integral is well-defined, thus $\int \d^3x \p^i F_{i0}^a=0$, and that $\int \d^3x\{Q_{BRST},D_0 \overline{c}^a\}$ is well defined too, we would have color confinement on physical states, due to $Q_{BRST}^2=0$. The big problem is of course to assure that the space integral $\int \d^3x\{Q_{BRST},D_0 \overline{c}^a\}$ is well-defined. It was discussed in  \cite{Kugo:1979gm,Kugo:1995km} that this is the case if
\begin{equation}
u(0)=-1,
\end{equation}
where $u(p^2)$ is the transverse part of a particular Green function, namely
\begin{eqnarray*}
\int \d^4x e^{ipx}\left\langle D_\mu c^a(x) D_\nu \overline{c}^b(0)\right\rangle=\delta^{ab}\left(\delta_{\mu\nu}-\frac{p_\nu p_\mu}{p^2}\right)u(p^2)-\frac{p_\mu p_\nu}{p^2}.
\end{eqnarray*}
An interesting relation now exists between the ghost propagator and this $u(p^2)$ in the specific case of the Landau gauge, first shown in \cite{Kugo:1995km} and confirmed in \cite{Kondo:2009ug},
\begin{eqnarray}\label{co2}
G^{ab}(p^2)=\left\langle c^a \overline{c}^b \right\rangle_p=\frac{\delta^{ab}}{p^2}\frac{1}{1+u(p^2)+p^2v(p^2)},
\end{eqnarray}
and recently rederived in a slightly different setting in the paper \cite{Aguilar:2009pp}. It is generally accepted that $v(0)=0$ \cite{Aguilar:2009pp}, and in a loop expansion, it is actually zero as one can learn from e.g. \cite{Ford:2009ar}, therefore
one notices that $u(0)=-1$ is realized when the ghost propagator is infrared enhanced, and vice versa. This explains the attention paid to the infrared behaviour of the ghost in the past, both from numerical and analytical viewpoint. We refer to e.g. \cite{Dudal:2008sp} for an extensive list of literature.

We can make a few comments. First of all, it was unclear at the time whether the Kugo-Ojima function $u(k^2)$ could be renormalized. In the Landau gauge, such a proof was presented in \cite{Dudal:2003dp,Grassi:2004yq}. However, it is clear that $u(0)=-1$ is not realized in perturbation theory, hence one needs to resort to nonperturbative calculations tools. One such example are lattice gauge studies, and most recent results agree on the fact that $u(0)=-1$ is not realized, either via direct studies or via the ghost propagator that appears to be not enhanced. Clearly, an important role in the analysis is played by the BRST charge, however it is unknown whether a well-defined BRST charge exists in the infrared. Moreover, the Kugo-Ojima analysis is purely based on the Faddeev-Popov action, and completely ignores the gauge copy problem.

\section{Combining Gribov-Zwanziger with Kugo-Ojima}
Having summarized in a nutshell the Gribov-Zwanziger approach and Kugo-Ojima results, it is instructive to compare both issues a bit. As already mentioned, the Gribov restriction suppresses the gluon in the infrared, which should make the first part in \eqref{color} well-defined. Simultaneously, the ghost is enhanced when the horizon condition is implemented, so apparently the Gribov-Zwanziger framework fulfills all the Kugo-Ojima requirements to ensure confinement. The correspondence becomes even more apparent when we compare \eqref{gh1} and \eqref{co2}, giving that $\sigma(0)=-u(0)$. The no-pole condition translates into $\sigma(0)=1$ at the quantum level, hence $u(0)=-1$. Recently, the correspondence between the nonlocal horizon function \eqref{gzz} and the ``pole function'' $\sigma(k^2)$ was also shown at higher orders \cite{Gomez:2009tj}.

The link between $u(p^2)$ and the horizon condition can also be formalized. Using the Ward identities of the Gribov-Zwanziger action, one shows that \cite{Zwanziger:1992qr}
\begin{eqnarray}\label{zwan}
\left\langle gf^{abc} A_\mu^a (\varphi+\overline{\varphi})_{\mu}^{bc} \right\rangle&=& 2\gamma^2\int \d^4x\left\langle D_\mu c^a(x) D_\mu \overline{c}^a(0)\right\rangle\\&=&-2\gamma^2(N^2-1)((d-1)u(0)-1).
\end{eqnarray}
Hence, implementing the horizon condition \eqref{hcond2} is equivalent with setting $u(0)=-1$.

We notice that the horizon condition can thus be connected to the correlator
\begin{eqnarray}
\left\langle D_\mu c^a(x) D_\mu \overline{c}^a(y)\right\rangle,
\end{eqnarray}
which can be done in a renormalizable setting. Other versions of the horizon function can be found in the literature \cite{Kondo:2009ug}, but to our understanding this would correspond to a connection with the correlator
\begin{eqnarray}
\left\langle (gf^{abc} A_\mu^b c^c(x)) (gf^{apq}A_\mu^p\overline{c}^q)(y)\right\rangle
\end{eqnarray}
which is however not renormalizable \cite{work}. We claim that the requirement of renormalizability fixes the ``choice'' in the horizon function in a unique way.

From the above considerations, it appears that there exists something like a GZKO framework, where everything fits wonderfully together. However, there is a big hole in this reasoning. As we already explained, the Gribov-Zwanziger restriction breaks the BRST, while the Kugo-Ojima analysis indispensably needs the BRST. This means that actually, in the Gribov-Zwanziger theory, the meaning of the Kugo-Ojima criterion is unclear.

A potential solution would be to construct a new BRST charge for the Gribov-Zwanziger theory. That this might be possible can be guessed from the fact that at the quadratic order,
\begin{eqnarray}
s S_{GZ}&=&s\left(\gamma^4 \int \d^4x A_\mu^a\frac{1}{\p^2}A_\mu^a\right)~\propto~ \gamma^4 \int \d^4x A_\mu \frac{1}{\p^2} \p_\mu c^a,
\end{eqnarray}
and performing a partial integration, it appears that this breaking could be absorbed into a redefinition of the $b^a$-field. A new BRST symmetry $\widetilde{s}$ for the complete theory was indeed found in \cite{Sorella:2009vt,Kondo:2009qz}, but it turns out to be nonlocal, as already explained on general grounds in \cite{Dudal:2008sp}. It remains unclear what a nonlocal BRST symmetry could be used for, it is even unclear what the corresponding charge would be. Evidently, a new BRST would also need a reworking of the Kugo-Ojima analysis, if possible of course.

Anyhow, till this moment, there is absolutely nothing proven about the nonperturbative unitarity of or confinement in the Yang-Mills theory in the Landau gauge, when relying on the Gribov-Zwanziger and/or Kugo-Ojima approach. All that one can say is that implementing the Gribov restriction leads to a violation of positivity in the spectral representation of the gluon propagator \cite{Dudal:2008sp}. This violation traces back the presence of complex poles in the propagator \eqref{gl1}, and as such the gluon cannot describe a physical particle. This violation of positivity has also been found in other approaches, like in \cite{Bowman:2007du,Alkofer:2003jj}.

\section{Imposing Kugo-Ojima as a boundary condition}
In this section, we wish to implement the Kugo-Ojima criterion as a boundary condition in the theory. This was worked out in \cite{Dudal:2009xh}, and it was inspired by works in which this is also done to select a specific solution. A very powerful tool to extract nonperturbative information from a quantum field theory is by solving their quantum equations of motion, being the Schwinger-Dyson equations. Although this is a horrendous task, much progress has been made in recent years, we refer to e.~g.~\cite{Binosi:2009qm,Fischer:2008uz,Boucaud:2008ky} for an overview of the literature and results. There exists whole classes of solutions for these equations, and one must impose certain conditions to discriminate between them. A possibility is to impose a condition on the ghost propagator in the deep infrared. One can imagine imposing an infrared enhanced ghost \cite{Fischer:2008uz}, which corresponds to the $a=0$ case of
\begin{equation}
    \left[p^2G(p^2)\right]^{-1}_{p^2=0}=a\geq0.
\end{equation}
It has been speculated that this choice of $a$ can be related to different nonperturbative implementations of the Landau gauge \cite{Maas:2009se}.

Since imposing a constraint on a propagator in an interacting quantum field theory corresponds to demanding something highly nontrivial at the quantum level, one can imagine that such a constraint might corrupt e.g.~some symmetry properties of the original action. We have explained in \cite{Dudal:2009xh} that for a consistent treatment, one would need to impose the constraint from the start, in order to have a decent way to discuss the symmetries of the constrained theory. It turns out that this is possible in case of the Kugo-Ojima criterion. More precisely, it is possible to impose that $u(0)=-1$, or thus that the ghost is infrared enhanced, and this in a way that is stable under quantum corrections. The latter is an important ingredient, as if this would not be possible, it would make no sense trying to enforce the theory's behaviour in such a way. In practice, we started from the following rather unconventional form of the Faddeev-Popov action,
\begin{eqnarray}
S_{YM+FP}' &=& S_{YM+FP}+
\int \d^{4}x\Bigl( \overline{\varphi }_{\mu }^{ac}\partial _{\nu}\left(\partial _{\nu }\varphi _{\mu }^{ac}+gf^{abm}A_{\nu }^{b}\varphi _{\mu}^{mc}\right)
-\overline{\omega }_{\mu }^{ac}\partial _{\nu }\left( \partial_{\nu }\omega _{\mu }^{ac}+gf^{abm}A_{\nu }^{b}\omega _{\mu }^{mc}\right)
\nonumber  \\
&&-g\left( \partial _{\nu }\overline{\omega }_{\mu}^{ac}\right) f^{abm}\left( D_{\nu }c\right) ^{b}\varphi _{\mu}^{mc}
\Bigr),
\end{eqnarray}
which is completely equivalent with the original one, \eqref{fp} (see Section 1). Next, we introduce a multiplier $\gamma^2$ as follows,
\begin{eqnarray}\label{actionko}
S_{YM+FP}'&\to& S_{KO}~\equiv~S_{YM+FP}'\nonumber\\ &-&\int \d^4x\Bigl(\gamma ^{2}g\left(f^{abc}A_{\mu }^{a}(\varphi _{\mu }^{bc}+\overline{\varphi }_{\mu }^{bc}) + \frac{4}{g}\left(N^{2}-1\right) \gamma^{2} \right)\Bigr),
\end{eqnarray}
and impose the condition $\frac{\p \Gamma}{\p \gamma^2}=0$, which enforces that
\begin{eqnarray}\label{constraint}
\left\langle gf^{abc} A_\mu^a (\varphi+\overline{\varphi})_{\mu}^{bc} \right\rangle&=& 2d(N^2-1)\gamma^2.
\end{eqnarray}
As we have already discussed before, this is in return sufficient to ensure the ghost enhancement, through $u(0)=-1$.

The reader shall of course observe the similarity between this result and the one concerning the Gribov-Zwanziger action. We want to stress here that currently, we are not saying \emph{anything} yet about copies or treatment thereof\footnote{But evidently, we rely on what we already learnt in previous subsections.}. What we are doing is imposing the constraint \eqref{constraint} on top of the normal Faddeev-Popov action. We then obtain an action that automatically implements the Kugo-Ojima criterion as a boundary condition, in way that is consistent at the quantum level, due to the renormalizability of the associated action \eqref{actionko}. We afterwards notice that the eventual action is indeed equivalent to the one constructed by Gribov and Zwanziger.

We conclude that imposing the Kugo-Ojima criterion on top of the Faddeev-Popov quantization softly breaks the BRST symmetry\footnote{It would be wonderful to find another way to impose the boundary, which preserves the BRST. Until now, we are unfortunately unaware of other consistent ways to do so.}. As such, the role of the condition $u(0)=-1$ in explaining confinement becomes again difficult to understand.

\section{Exploring the Gribov-Zwanziger quantum dynamics}
In this concluding section, we shall spend a few words on the quantum dynamics of the Gribov-Zwanziger theory. The motivation arises from the comparison with lattice results, which essentially come to the agreement that the gluon propagator is suppressed in the infrared, but it does not vanish at zero momentum, while the ghost propagator does not seem to be enhanced \cite{Cucchieri:2008mv} (see however e.g. \cite{Oliveira:2008uf,Maas:2008ri,Sternbeck:2008mv} for other views).

It is perhaps important to notice that the restriction to the Gribov region $\Omega$ in fact corresponds to selecting the minima along the gauge orbit of the functional $\int \d^4x A_\mu^2$. The problem of residual copies is caused by the existence of multiple local minima. One should resort to finding the absolute minima \cite{vanBaal:1991zw}. In the continuum, it has been argued that it would be sufficient to restrict to $\Omega$ when calculating expectation values \cite{Zwanziger:2003cf}. In the continuum, the restriction to $\Omega$ is done via the Gribov-Zwanziger action \eqref{gz1}. It is however not possible to simulate with the Gribov-Zwanziger action on a lattice. Numerically, one rather selects the gauge configuration that corresponds to the ``best'' minimum each time. It is always assumed that in the infinite volume limit, the lattice and the continuum version are essentially doing the same.

This being said, in the past, using relatively small lattices, there seemed to be agreement between the continuum predictions and the numerical estimates. The situation changed when the papers \cite{Cucchieri:2007md,Bogolubsky:2007ud} appeared, where significantly larger lattices were invoked. Improved Schwinger-Dyson studies were able to reproduce these lattice results \cite{Binosi:2009qm,Fischer:2008uz,Boucaud:2008ky}. But the Gribov-Zwanziger framework seemed to miss something essential. In  \cite{Dudal:2008sp,Dudal:2007cw}, we paid attention to other potentially important nonperturbative effects in the Gribov-Zwanziger theory, which could also influence the propagators.

In particular, we focused on the dynamics of the additional fields,$(\overline{\varphi}_\mu^{ac}$, $%
\varphi_\mu^{ac} $, $\overline{\omega}_\mu^{ac}$,$\omega_\mu^{ac}$), which had to be introduced to consistently treat the theory. In a sense, these fields describe the influence of the Gribov horizon. Evidently, they will develop their own quantum dynamics, which can in return influence the conventional gluon and ghost sector, as these are nontrivially coupled to the ``auxiliary'' sector. In \cite{Dudal:2008sp}, we presented evidence for the existence of the dimension 2 condensate, $\braket{\overline{\varphi}_{\mu}^{ac}\varphi_{\mu}^{ac}-\overline{\omega}_{%
\mu}^{ac}\omega_{\mu}^{ac}}$. This operator is a very natural object to study, as the Gribov-Zwanziger theory already has an inherent mass scale in it, being the Gribov parameter $\gamma^2$, and the horizon condition exactly corresponds to a $d=2$ condensate, \eqref{hcond2}. Moreover, the operator is also compatible with the renormalizability \cite{Dudal:2007cw,Dudal:2008sp}. To get a taste of the importance of such a condensate, we studied it in \cite{Dudal:2008sp} using variational perturbation theory. Doing so, we found a gluon propagator of the type
\begin{equation}\label{gluon}
    D(p^2)=\frac{p^2+M^2}{p^4+M^2p^2+\lambda^4},
\end{equation}
while for the ghost, we obtained
\begin{equation}
    G(p^2)\sim\frac{1}{p^2}\,,\qquad\textrm{for } p^2\sim 0.
\end{equation}
The mass parameter $M^2$ corresponds to the condensate $\braket{\overline{\varphi}_{\mu}^{ac}\varphi_{\mu}^{ac}-\overline{\omega}_{%
\mu}^{ac}\omega_{\mu}^{ac}}$, and it is fixed in a self-consistent way through the variational perturbative approach \cite{Dudal:2008sp}. Both predictions of the so-called Refined Gribov-Zwanziger framework \cite{Dudal:2008sp} are now in compliance with the lattice data.

We can reexpress the gluon propagator \eqref{gluon} as
\begin{equation}\label{gluon}
    D(p^2)=\frac{1}{p^2+ m^2(p^2)}\,,\qquad\textrm{with } m^2(p^2)=\frac{\lambda^4}{p^2+M^2}.
\end{equation}
This kind of ``effective gluon mass'' $m^2(p^2)$, which vanishes in the ultraviolet, has also been found in Schwinger-Dyson studies \cite{Binosi:2009qm,Cornwall:1981zr}, and finds phenomenological application, see e.g. \cite{Mathieu:2009cc}.

We are currently studying the issue of $d=2$ condensates in the Gribov-Zwanziger theory more thoroughly, by using an effective potential approach for local composite operators \cite{work}. We are not only including the aforementioned one, but also other ones like $\braket{\varphi_{\mu}^{ac}\varphi_{\mu}^{ac}}$, see also \cite{Sorella:2009vt}. If a nonvanishing vacuum expectation values for such condensates is dynamically favoured as the vacuum energy is lowered, then this would be an excellent illustration of the fact that the dynamics of the Gribov-Zwanziger theory is very rich, and needs to be taken into proper account to obtain reliable estimates for e.g. the propagators. Before finishing, it is perhaps important to point out that none of these condensates are free to choose mass parameters \cite{Fischer:2008uz}, each one of them is eventually expressed in terms of $\Lambda_{QCD}$ by dimensional transmutation\footnote{Notice that the horizon condition also fixes the Gribov parameter $\gamma^2$ in terms of  $\Lambda_{QCD}$ \cite{Dudal:2005na}.}.

An issue we did not touch in this proceeding is the question what the physical operators might be in the Gribov-Zwanziger theory, since we do not longer have the BRST symmetry. We refer to \cite{Dudal:2009zh,Vandersickel:2009uv,work2} for a discussion thereof.

\section*{Acknowledgments.}
We thank the organizers for the kind invitation to present our work. We are also grateful to all participants with whom we shared the pleasure of discussing, in particular O.~Oliveira. D.~Dudal and N.~Vandersickel are  supported by the Research Foundation-Flanders (FWO). The Conselho Nacional de
 Desenvolvimento Cient\'{\i}fico e Tecnol\'{o}gico (CNPq-Brazil), the Faperj,
 Funda{\c{c}}{\~{a}}o de Amparo {\`{a}} Pesquisa do Estado
 do Rio de Janeiro, the SR2-UERJ and the Coordena{\c{c}}{\~{a}}o de
 Aperfei{\c{c}}oamento de Pessoal de N{\'{\i}}vel Superior (CAPES),
 the CLAF, Centro Latino-Americano de F{\'\i}sica, are gratefully acknowledged for financial support.

\end{document}